\documentclass[twocolumn,showpacs,preprintnumbers,amsmath,amssymb]{revtex4}

\usepackage{graphicx}
\usepackage{dcolumn}
\usepackage{bm}
\usepackage{color}
\def\t2{\mbox{  }}
\def\rst1{\mbox{ }}

\begin{document}

\preprint{LPT-ORSAY : 10-16}

\title{The Metropolis Monte Carlo Finite Element Algorithm for Electrostatic Interactions.}

\author{Martial MAZARS}
\email{Martial.Mazars@th.u-psud.fr}
\affiliation{Laboratoire de Physique Th\'eorique (UMR 8627),\\
\small Universit\'e de Paris Sud XI, B\^atiment 210, 91405 Orsay Cedex,
FRANCE}%

\date{\today}

\begin{abstract}
The Metropolis Monte Carlo algorithm with the Finite Element method applied to compute electrostatic interaction energy between charge densities is described in this work. By using the Finite Element method to integrate numerically the Poisson's equation, it is shown that the computing time to obtain the acceptance probability of an elementary trial move does not, in principle, depend on the number of charged particles present in the system. 
\end{abstract}

\maketitle
Long ranged electrostatic interactions are particularly difficult to take into account correctly in computer simulation of condensed matter. At present and to this aim, the most adequate methods, in accuracy and efficiency, are the Ewald summations \cite{deLeeuw:80,Darden:93,Deserno:98}. In classical electrodynamics, the long ranged feature of electrostatic interactions is included in the structure of the local partial differential equation for the electrostatic field (Poisson, Laplace, etc.) and the particular analytical form of these interactions is governed by the boundary conditions \cite{Jackson:75}. In Ewald sums, one uses (partial) periodic boundary conditions \cite{Smith:08,Arnold:02} and a macroscopic boundary condition (surface term) to obtain univocally energy of systems \cite{Smith:08,Arnold:02,deLeeuw:81,Smith:94,Fraser:96,Herce:07} ; then the energy computed likewise is used in Metropolis algorithm to sample the phase space of the system \cite{Metropolis:53} in Monte Carlo simulations or, in Molecular dynamics, one computes forces in similar ways.\\
The Finite Element method allows to obtain numerical approximations of partial differential equations with boundary conditions \cite{Hughes:00,Braess:01,Ramdas:02} ; applied to Poisson's equation, they define Poisson solvers that may be used to compute electrostatic fields and energy. Such Poisson solvers have been used in {\it ab initio} electronic structure calculations in solids \cite{Pask:01,Pask:05}. The Finite Element method combined with Ewald sums has also been proposed recently for the electrostatic interactions in slablike geometry \cite{Alireza:07}.\\
In this letter, we show how to use the Finite Element method to obtain acceptance probability of trial moves in Metropolis Monte Carlo algorithms in systems with electrostatic interactions. The work presented here aims to describe theoretically the algorithm ; to be considered as an efficient alternative to Ewald methods, detailed comparisons are still to be made. The main result of the present work is the derivation of the electrostatic energy of a trial configuration from the energy of the old configuration and the variations of few shape functions that gives locally an estimate of the electrostatic field (Eq.\ref{FEM21}).\\
For a charge density distribution $\rho(\bm{x})$ on a domain $\Omega$ of the space, the electrostatic potential is given by the Poisson's equation as 
\begin{equation}
\label{FEM1}
\nabla^2 V = -4\pi\rho
\end{equation} 
On boundaries $\Gamma$ of $\Omega$ some conditions must be fulfilled by the potential to obtain an unique solution. These boundary conditions can be chosen by specificying the value of the electrostatic potential on $\Gamma$ (Dirichlet problem) or the value of the electric field (Neumann problem) \cite{Jackson:75}. One can also show that Eq.(\ref{FEM1}) has an unique solution for mixed boundary conditions : taking Dirichlet boundary conditions on $\Gamma_{D}$ and Neumann boundary conditions on $\Gamma_{N}$ with $\Gamma=\Gamma_{D}\cup\Gamma_{N}$. For mixed boundaries conditions, the electrostatic potential is constrained by  
\begin{equation}
\label{FEM2}
\forall \bm{x}\in \Gamma_{D}, \rst1\t2 V(\bm{x})=V_{0}(\bm{x})
\end{equation}
and
\begin{equation}
\label{FEM3}
\forall \bm{x}\in \Gamma_{N}, \rst1\t2 \frac{\partial V}{\partial n} \hat{\bm{n}}(\bm{x})=-\bm{E}_{0}(\bm{x})
\end{equation}
where $V_{0}$ and $\bm{E}_{0}$ are known functions and they are called boundary conditions on $\Gamma$. Eqs. (\ref{FEM1}-\ref{FEM3}) define the so-called strong form of a boundary-value problem in the finite element method \cite{Hughes:00}. The Finite Element method starts by deriving a weak formulation of the boundary value problem ; strong and weak formulations of the boundary-value problem are fully equivalent. The weak formulation for Poisson's equation reads as follows : for any weighting functions (or variations) $w\in \Xi$, find the function $V\in \Upsilon$ such as    
\begin{equation}
\label{FEM4}
\int_{\Omega}\bm{\nabla}w.\bm{\nabla}V\rst1 d\Omega = 4\pi\int_{\Omega}w\rho\rst1 d\Omega -\int_{\Gamma_{N}}w\rst1 \hat{\bm{n}}.\bm{E}_{0}\rst1 d\Gamma
\end{equation}
where the set $\Upsilon$ of trial functions and the set $\Xi$ of weighting functions are defined respectively by
\begin{equation}
\left \{\begin{array}{lcl}
\label{FEM5}
\Upsilon & =  & \{v\in H^{1}(\Omega)\rst1 :\rst1 \forall \bm{x}\in\Gamma_{D}, v(\bm{x})=V_{0}(\bm{x}) \}\\
&&\\
\Xi & = & \{w\in H^{1}(\Omega)\rst1 :\rst1\forall \bm{x}\in\Gamma_{D}, w(\bm{x})=0 \}
\end{array}
\right.
\end{equation}
with  $H^{1}(\Omega)$, the Sobolev space of degree one \cite{Hughes:00,Ramdas:02}.\\ 
The boundary conditions given by Eq.(\ref{FEM2}) are called essential boundary conditions, they are explicitly involved in the definition of $\Upsilon$, while the boundary conditions given by Eq.(\ref{FEM3}) are called natural boundary conditions and they are incoporated into Eq.(\ref{FEM4}) by using the Green first identity in the derivation of the weak formulation. If the essential boundary conditions have finite values on each point on $\Gamma_{D}$, then the weighting functions $w$ are required to satisfy homogeneous counterparts of the essential boundary conditions \cite{Hughes:00}, as defined by the set $\Xi$, Eq.(\ref{FEM5}).\\
Having the weak form of a boundary-value problem, the Finite Element method then provides an approximation for the function $V$. The first step is to construct finite-dimensional representations $\Upsilon^{h}$ and $\Xi^{h}$ as approximations for $\Upsilon$ and $\Xi$ ; we require that $\Upsilon^{h}\subset\Upsilon$ and $\Xi^{h}\subset\Xi$. Both sets $\Upsilon^{h}$ and $\Xi^{h}$ are associated to a mesh that divides $\Omega$ into domain elements whose characteristic length or volume is represented by $h$. In the following, we restrict the discussion to Galerkin's approximation scheme for the Finite Element Method of $\mathcal{C}^0$-class (for instance see Chap.2 of ref.\cite{Braess:01} and refs.\cite{Hughes:00,Ramdas:02}).\\
On element domains, nodal points and shape functions, associated to each nodal point, are defined and numbered (locally and globally). We use the following notations : the position of a nodal point numbered $i$ is $\bm{x}_i$ ; the set of all nodal points is $\bar{\omega}$ ; the set of all nodal points lying on $\Gamma_{D}$ is called $\gamma_{D}$ ; the set of nodal points not lying on $\Gamma_{D}$ is called $\omega$ and the function $\phi_{i}$ is the shape function associated to the nodal point numbered $i$. We note also $\mid\omega\mid$, the number of nodal points in $\omega$. We require $\phi_{i}(\bm{x}_j)=\delta_{ij}$, where $\delta_{ij}$ is the Kronecker symbol. The nodal points are particular locations in $\Omega$ where the approximate solution given by the Finite Element method is the most accurate.\\  
For $w^{h}\in\Xi^{h}$, the method assumes that it can be written as 
\begin{equation}
\label{FEM7}
w^{h}(\bm{x})=\sum_{i\in\bar{\omega}}c_{i}\phi_{i}(\bm{x})
\end{equation} 
where $c_{i}$ is a constant, i.e. the value of $w^{h}$ at node $i$. For $w^{h}\in\Xi^{h}$ and $i\in\Gamma_{D}$, we have $c_{i}=0$. The approximate solution of the weak form admits the representation 
\begin{equation}
\label{FEM8}
V^{h}=v^{h}+V_{0}^{h}  
\end{equation}    
where $v^{h}\in\Xi^{h}$ and $V_{0}^{h}$ result in satisfaction of the boundary condition given by Eq.(\ref{FEM2}), these functions are expanded according to
\begin{equation}
\label{FEM9}
v^{h}(\bm{x})=\sum_{i\in\omega}v_{i}\phi_{i}(\bm{x})
\end{equation}
where the $v_{i}$ are the unknowns values of $v^{h}$ at each node $i$ and 
\begin{equation}
\label{FEM10}
V_{0}^{h}(\bm{x})=\sum_{i\in\gamma_{D}}V_{0}(\bm{x}_i)\phi_{i}(\bm{x})
\end{equation}
With this representation, the integral equation Eq.(\ref{FEM4}) becomes an algebraic equation with $\mid\omega\mid$ unknowns. One may easily obtain 
\begin{equation}
\label{FEM11} 
\begin{array}{ll}
\displaystyle\sum_{j\in\omega}a(\phi_i,\phi_j)v_j = 4\pi (\phi_i,\rho)&\displaystyle-( \phi_i,\hat{\bm{n}}.\bm{E}_{0})_{\Gamma_{N}}\\
&\displaystyle -\sum_{k\in\gamma_{D}}a(\phi_i,\phi_k)V_{0}(\bm{x}_k)
\end{array}
\end{equation}
where we have set
\begin{equation}
\label{FEM12}
\left \{\begin{array}{lcl}
\displaystyle A_{ij}=a(\phi_i,\phi_j)&=&\displaystyle\int_{\Omega}\bm{\nabla}\phi_i .\bm{\nabla}\phi_j\rst1 d\Omega\\
&&\\
\displaystyle (\phi_i,\rho)&=&\displaystyle \int_{\Omega}\phi_i\rho\rst1 d\Omega\\
&&\\
\displaystyle(\phi_i,\hat{\bm{n}}.\bm{E}_{0})_{\Gamma_{N}}&=&\displaystyle\int_{\Gamma_{N}}\phi_i\rst1 \hat{\bm{n}}.\bm{E}_{0}\rst1 d\Gamma
\end{array}
\right.
\end{equation}
Eq.(\ref{FEM11}) is the matrix form of the weak formulation of the boundary-value problem in the Galerkin approximation scheme. An estimation of $V(\bm{x})$ is therefore obtained as in Eq.(\ref{FEM8}) by solving the linear algebraic equations given by Eq.(\ref{FEM11}) to obtain the constants $v_j$. With convenient choices of nodes and shape functions, the matrix in the left handed side of Eq.(\ref{FEM11}) is sparse.\\
When the boundary conditions used are periodic, for instance in {\it ab initio\/} solid-state electronic structure computations \cite{Pask:01,Pask:05} or in Molecular simulations, for all $\bm{x}$ on the boundaries ($\Gamma$) of $\Omega$, the electrostatic potential must fulfill   
\begin{equation}
\left \{\begin{array}{lcl}
\label{FEM13}
V(\bm{x})&=&V(\bm{x}+\bm{L}_{\alpha})\\
&&\\
\hat{\bm{n}}(\bm{x}).\bm{\nabla}V(\bm{x})&=&-\hat{\bm{n}}(\bm{x}+\bm{L}_{\alpha}).\bm{\nabla}V(\bm{x}+\bm{L}_{\alpha})
\end{array}
\right.
\end{equation}
Where the domain $\Omega$ is chosen as a parallelepipedic cell (orthorhombic or triclinic), $\hat{\bm{n}}$ the outward normal unit vector and $\bm{L}_{\alpha}$ are the lattice vectors associated with the periodic boundaries conditions. These boundary conditions are easily fulfilled with a clever global numbering of nodal points (cf. ref.\cite{Pask:01,Pask:05}). Partial periodic boundary conditions are used to represent quasi-one and quasi-two dimensional systems (nanotubes, surfaces, interfaces, biological membranes, etc.) ; for these systems, Eqs.(\ref{FEM13}) are applied for directions having spatial periodicities, while Eqs.(\ref{FEM2},\ref{FEM3}) for other directions. Thus, the method seems particularly well adapted to the representation of systems with partial periodic boundary conditions.\\
For a set of point charges $\{q_{\alpha}\}_{\alpha=1,N}$ located at $\{\bm{x}_{\alpha}\}_{\alpha=1,N}$, the charge density is given by 
\begin{equation}
\label{FEM15}
\displaystyle \rho(\bm{x})=\sum_{\alpha=1}^{N}q_{\alpha}\delta(\bm{x}-\bm{x}_{\alpha})
\end{equation}
then, the electrostatic potential is given by Eqs.(\ref{FEM8},\ref{FEM9}) with $v_j$ the solutions of Eq.(\ref{FEM11}). More precisely,
\begin{equation}
\label{FEM16}
\begin{array}{ll}
&\displaystyle v_j(\{\bm{x}_{\alpha}\}) \displaystyle =4\pi \sum_{\alpha=1}^N q_{\alpha}\sum_{i\in\omega} (A^{-1})_{ji}\phi_i(\bm{x}_{\alpha})\\
&\\
&\displaystyle -\sum_{i\in\omega} (A^{-1})_{ji} [ (\phi_i,\hat{\bm{n}}.\bm{E}_{0})_{\Gamma_{N}}+\sum_{k\in\gamma_{D}} a(\phi_i,\phi_k)V_{0}(\bm{x}_k) ]
\end{array}
\end{equation}
where $A^{-1}$ stands for the inverse matrix of matrix $A$ defined in Eq.(\ref{FEM12}). The approximation for the electric field is $\bm{E}^h=-\bm{\nabla} V^{h}$ and, for natural boundary conditions (Neumann), the energy of the configuration is given by
\begin{equation}
\label{FEM17}
\begin{array}{ll}
\displaystyle W &\displaystyle=\frac{1}{8\pi}\int_{\Omega}\mid \bm{E}^h \mid^2 d\bm{x}\\
&\\
&\displaystyle = \frac{1}{8\pi}\sum_{i\in\omega}\sum_{j\in\omega}v_i(\{\bm{x}_{\alpha}\}) A_{ij}v_j(\{\bm{x}_{\alpha}\})
\end{array}
\end{equation}
An elementary trial move for the particle numbered $a$ is $\bm{x}_{a}\rightarrow\bm{x}'_{a}$ ; from this elementary move the trial configuration of the set of charges is $\{\bm{x}_{\alpha}\}'$. Then, the trial configuration is accepted or rejected according to the Metropolis algorithm. From Eq.(\ref{FEM11}) and (\ref{FEM12}), the electrostatic potential for the trial configuration is given by the electrostatic potential of the old configuration as  
\begin{equation}
\label{FEM18}
\displaystyle v_i(\{\bm{x}_{\alpha}\}')=v_i(\{\bm{x}_{\alpha}\})+4\pi q_a\sum_{j\in\omega} (A^{-1})_{ij}\Delta\phi_j(a)
\end{equation}
where 
\begin{equation}
\label{FEM19}
\Delta\phi_j(a)=\phi_j(\bm{x}'_a)-\phi_j(\bm{x}_a)
\end{equation}
More precisely, the approximate solution of the electrostatic potential for the trial configuration is thus given by 
\begin{equation}
\label{FEM20}
\begin{array}{ll}
V^{h}(\{\bm{x}_{\alpha}\}',\bm{x}) &\displaystyle = V^{h}(\{\bm{x}_{\alpha}\},\bm{x})\\
&\\
&\displaystyle +4\pi q_a \sum_{i\in\omega} \phi_i(\bm{x})\sum_{j\in\omega} (A^{-1})_{ij}\Delta\phi_j(a)
\end{array}
\end{equation}
and, from Eqs.(\ref{FEM17}) and (\ref{FEM18}), we obtain
\begin{equation}
\label{FEM21}
\begin{array}{ll}
\displaystyle W' &\displaystyle = W+ q_a\sum_{i\in\omega}v_i(\{\bm{x}_{\alpha}\})\Delta\phi_i(a)\\
&\\
&\displaystyle +2\pi q_a^2\sum_{i\in\omega}\sum_{k\in\omega}\Delta\phi_i(a)(A^{-1})_{ik}\Delta\phi_k(a)
\end{array}
\end{equation}
This last equation is the main result of this letter. From Eq.(\ref{FEM21}), we may accept or reject the trial move by using the standard Metropolis scheme with the probability $\mbox{min}(1,\exp(-\beta(W'-W)))$. \\
If the trial move is accepted, then the new electrostatic potential is obtained by updating $v_i(\{\bm{x}_{\alpha}\})$ using Eq.(\ref{FEM18}) ; thus an another trial move may be done from the new configuration. If the trial move changes the location of more than one charge, as it is the case for polyatomic molecules such as water or even more complicated molecules as DNA, proteins, RNA, etc., then, since equations are linear, Eq.(\ref{FEM18}) and (\ref{FEM21}) must include a summation over charges that are involved in the trial move.\\
In principle, such algorithm is very efficient since each charge is located in only one element domain and the shape function $\phi_i$ is different from $0$ only in domain elements that contain the nodal point $\bm{x}_i$ associated with the shape function $\phi_i$. If there are $n$ nodal points in each domain element, then there are at most $2n$ values  $\Delta\phi_j(a)$ defined by Eq.(\ref{FEM19}) that are non zero and that contribute to Eq.(\ref{FEM21}). Once all $\Delta\phi_j(a)$ are computed, to evaluate $W'$ from $W$ the number of additions and products needed scale as $4n^2$.\\
For concrete implementations of this method in a standard NVT Monte Carlo simulations, one may proceed as follows :
\begin{itemize}
\item[(a)] The domain $\Omega$ is defined as the simulation box, it is divided into domain elements ; locations of nodal points in each elements and shape functions for each nodal point are defined. Boundary conditions Eqs.(\ref{FEM2},\ref{FEM3}) and Eq.(\ref{FEM13}) are applied. (This step defines the matrix $A$.)
\item[(b)] The inverse of matrix $A$ is computed and stored.
\item[(c)] An initial configuration for charges (or charge density) is build. (After this step, the charge density is defined and with step (a) all contributions in the right handed side of Eq.(\ref{FEM11}) are defined.)
\item[(d)] The algebraic equation (\ref{FEM11}) is solved by using the matrix $A^{-1}$. (This step gives an approximation of the electrostatic potential as Eq.(\ref{FEM8}).)
\item[(e)] Electrostatic energy of initial configuration is computed by using Eq.(\ref{FEM17}). (The cpu-time of this step scales as $N^2$ where $N$ is the number of charges in the domain $\Omega$.)
\item[(f)] Trial moves are performed and they are accepted or rejected according to the Metropolis algorithm by using Eq.(\ref{FEM21}) ; averages are accumulated. (According Eq.(\ref{FEM21}), only matrix $A^{-1}$, computed in step (b), is needed to compute the energy of the trial configuration from the energy of the old configuration ; thus, the cpu-time needed for each trial moves does not depend on $N$.)
\end{itemize}
This method may also be applied to Monte Carlo simulations that sample the shape of the simulation box (NPT, etc.). In these implementations, steps (a) to (e) have to be done for each trial move  of the box shape.\\ 
In its principles, the Monte Carlo Finite Element algorithm presented in this letter has several benefits in comparison to others algorithms for electrostatic long ranged interactions as Ewald methods, etc.\\
First, and as the main benefit, the algorithm is local. The summations over the pairs of particles is done only one time in the whole computation, at step (e). As a consequence of Eq.(\ref{FEM21}), the $cpu$-time needed to compute the acceptance probability of a trial move does not depend on the number of particles in the system ; this makes the method far more efficient than Ewald methods.\\
By comparison with mesh based Ewald methods (PME, SPME, P$^3$M, etc.) \cite{Darden:93,Deserno:98}, it is not necessary to assign charge to mesh points : each charge keeps its own identity.\\
By comparison with auxiliary field local algorithm developed in refs.\cite{Maggs:02}, one does not have to introduce a gauge invariant constrained auxiliary field in the Hamiltonian of the system to compute energy.\\
Finally, another important benefit of this algorithm is practical. Finite Element Method is one of the most used method to solve numerically partial differential equations, therefore there are a lot of very efficient codes that already exists.\\ 
Nevertheless, the algorithm presented in the present letter can not yet be considered as an alternative to existing methods ; to be so, the efficiency and accuracy of the Monte Carlo Finite Element algorithm has to be compared to the efficiency and accuracy of Ewald methods. Such a comparison is still to be made, preferencially on a simple system.\\
There are mainly two points that strongly affect the accuracy and efficiency of the algorithm. First, the number of domain elements and the number of nodal points are critical for the convergence rate of approximate solutions $V^{h}$ to electrostatic potential $V$. These two numbers define the size of matrix $A$ and $A^{-1}$ : $\mid\omega\mid\times \mid\omega\mid$. If the value $\mid\omega\mid$ is too large to obtain an accuracy similar to the one achieved in Ewald methods, then the efficiency may drastically fall because of processors memory access problems. Second, it is necessary to compute energy of the system as in Eq.(\ref{FEM17}) and not as 
\begin{equation}
\label{FEM22}
W=\frac{1}{2}\int_{\Omega}\rho(\bm{x})V^h(\bm{x})d\bm{x}
\end{equation}
Otherwise, contributions to energy from self interactions are lost (cf. Chap.1 of ref.\cite{Jackson:75}) and comparisons with Ewald methods become difficult to achieve. Clearly, this implies that Finite Elements of $\mathcal{C}^1$-class have to be used to ensure the continuity of the electric field across boundaries of domain elements ; to apply the method to molecular dynamic simulations this requirement  is essential too.\\
Equation (\ref{FEM9}), that defines the unknowns $v_i$, corresponds to element of $\mathcal{C}^0$-class : the approximation $V^{h}$ is continuous on $\Omega$, but derivatives of $V^{h}$ are not continuous across boundaries of domain elements. For elements of $\mathcal{C}^1$-class, the equation that defines the unknowns is written as
\begin{equation}
\label{FEM23}
v^{h}(\bm{x})=\sum_{i\in\omega}\left( v_{i}\phi_{i}(\bm{x}) + v_i^{(\alpha)}\phi_{i}^{(\alpha)}(\bm{x}) \right)
\end{equation}
with the same notations as in equation (\ref{FEM9}) and $\phi_{i}^{(\alpha)}$ the shape functions for partial derivatives (they verify : $\phi_{i}^{(\alpha)}(\bm{x}_j)=0$ and $\partial_{\alpha}\phi_{i}^{(\alpha)}(\bm{x}_j)=\delta_{ij}$) and $v_i^{(\alpha)}$ the unknowns for derivatives. In one hand, because of equation (\ref{FEM23}), the number of unknowns by domain elements of $\mathcal{C}^1$-class is increased by comparison to elements of $\mathcal{C}^0$-class, thus is the size of matrix $A$ and $A^{-1}$. On the other hand, the convergence rate of Finite Element Method of $\mathcal{C}^1$-class is extremely better than $\mathcal{C}^0$-class. Therefore some optimal choices for shape functions, structure of domain elements, etc. are to be made.\\

\vfill


\end{document}